\documentclass[11pt,a4paper]{article}
\usepackage{jheppub_kim}
\usepackage{rotating}  
\usepackage{graphicx,epsfig}
\usepackage{epstopdf}
\usepackage{amsmath}
\usepackage {amssymb}
\usepackage{subfigure}

\usepackage{relsize}

\providecommand{\U}[1]{\protect\rule{.1in}{.1in}}

\newcommand{\newc}{\newcommand}

\newc{\be}{\begin{equation}}
\newc{\ee}{\end{equation}}

\newc{\ba}{\begin{eqnarray}}
\newc{\ea}{\end{eqnarray}}
\newc{\bea}{\begin{eqnarray*}}
\newc{\eea}{\end{eqnarray*}}
\newc{\D}{\partial}
\newc{\ie}{{\it i.e.} }
\newc{\eg}{{\it e.g.} }
\newc{\etc}{{\it etc.} }
\newc{\etal}{{\it et al.}}
\newc{\lcdm}{$\Lambda$CDM }

\newc{\ra}{\Rightarrow}

\title{Inflation with non-canonical scalar fields revisited }

\author[a]{Smaragda Lola}

\author[a]{Andreas Lymperis}

\author[b,c,d]{Emmanuel N. Saridakis}

\affiliation[a]{Department of Physics, University of Patras, 26500 Patras, 
Greece}

\affiliation[b]{National Observatory of Athens, Lofos Nymfon, 11852 Athens, 
Greece}

\affiliation[c]{Department of Physics, National Technical University of Athens, 
Zografou
Campus GR 157 73, Athens, Greece}
\affiliation[d]{Department of Astronomy, School of Physical Sciences, 
University of Science and Technology of China, Hefei 230026, P.R. China}

\emailAdd{magda.lola@upatras.gr}
\emailAdd{alymperis@upatras.gr}
\emailAdd{msaridak@phys.uoa.gr}

\abstract{ 
We revisit inflation with non-canonical scalar fields by 
applying deformed-steepness exponential potentials. 
We show that the resulting scenario can lead to inflationary observables, and 
in particular to  scalar spectral index   and   tensor-to-scalar ratio, in 
remarkable agreement with observations. Additionally, a significant 
advantage of the scenario  is that the required parameter values, such 
as the   non-canonicality exponent and scale, as well as the potential exponent 
and scale, do not need to acquire unnatural values and hence can accept a 
theoretical justification. Hence, we obtain a   significant improvement with 
respect to alternative schemes, and we present distinct correlations 
between the model parameters that  better fit the data, which
can be tested in future probes.  This combination of observational 
efficiency and theoretical justification makes the scenario at hand a good 
candidate for the description of inflation.  
}
\keywords{Inflation, non-canonical scalar fields, deformed-steepness potentials}

\begin{document}
\maketitle

\section{Introduction}

Inflation 
 is now  a crucial part of the Standard Model of Cosmology 
\cite{Starobinsky:1980te,Kazanas:1980tx,Sato:1980yn,Guth:1980zm,Linde:1981mu}. 
Its solution to the horizon and flatness
problems, together with the predictions for  an almost scale invariant 
perturbation spectral index,
have been confirmed by measurements of the  cosmic microwave
background (CMB)  radiation.
Nevertheless, the specific mechanism that triggers the inflationary epoch  is 
one  of the most outstanding issues in contemporary particle  physics and 
cosmology. As a result, the building of
theoretical models that explain this early
accelerating expansion of the universe   has   
exploded in
recent years. The first main class of mechanisms that can lead 
to successful inflation is based on the introduction of a scalar
field, while the second main class is obtained through gravitational modifications 
(for 
 reviews see 
\cite{Olive:1989nu, Lyth:1998xn,Bartolo:2004if,Nojiri:2010wj,Martin:2013tda}). 
Consequently, inflation-related observations 
have provided significant insight to both modified gravity 
\cite{Appleby:2009uf,Nojiri:2003ft,Carter:2005fu,Ferraro:2006jd,Germani:2010gm,
Sebastiani:2013eqa}, as well as to 
particle physics model building.
The literature on the latter is very extensive,
particularly within the framework of
supersymmetry \cite{Ellis:1982ed, Ellis:1982yb, Dvali:1994ms, Ross:1995dq},  
supergravity \cite{Nanopoulos:1982bv, Stewart:1994ts, Halyo:1996pp, 
Linde:1997sj}, theories of
extra dimensions such as superstring and brane theories
\cite{Shafi:1986vv, Antoniadis:1988aa, Dvali:1998pa},
and   technicolor too \cite{Adams:1992bn}.
Detailed lists of references on different theoretical constructions can be found in \cite{Olive:1989nu,
Lyth:1998xn,Martin:2013tda}.

In trying to understand the above issues (often in the framework of a
single theory) several problems have been
encountered, including fine-tuning issues (tiny dimensionless constants) and large
predictions for tensor fluctuations. In this respect, theories of
scalar fields with 
non-canonical kinetic terms, as expected in supergravity and superstring
theories, including the $k$-inflation subclass
\cite{ArmendarizPicon:1999rj,Garriga:1999vw,Mukhanov:2005bu}, 
were found to have significant advantages.
These theories arise commonly in the framework of supergravity and
string compactifications, which typically contain a
large number of light scalar fields $X$ (moduli), whose dynamics are
governed by a non-trivial moduli space metric $G_{ij}$.
As long as the moduli space metric is not flat, we generically expect
non-canonical kinetic terms.
Such effects could, but need not, be suppressed by the high scale of
the corresponding
Ultra-Violet physics (e.g. moduli masses, string scale), but they can still have
significant cosmological consequences
through the dynamics of the dilaton and moduli fields.

Among their many advantages, non-canonical scalars satisfy in a more efficient 
way the
slow-roll conditions of inflation,   since the
additional effective friction terms in the equations of motion of the
inflaton  slow down the scalar field for
potentials which would otherwise be too steep. Hence, the resulting 
tensor-to-scalar ratio is  significantly reduced 
\cite{Barenboim:2007ii,Tzirakis:2008qy,Franche:2009gk,Franche:2010yj,
Unnikrishnan:2012zu,Gwyn:2012ey,Easson:2012id,Zhang:2014dja,Gwyn:2014wna, 
Hossain:2014xha,
Rezazadeh:2014fwa,Cespedes:2015jga,Stein:2016jja, 
Dimopoulos:2017zvq,Mohammadi:2018wfk,Naderi:2018kre, 
Kamenshchik:2018sig,Do:2020ler}.
Moreover, models with non-canonical kinetic terms often allow for the kinetic 
term to
play the role of dark matter and the potential terms to generate dark
energy and inflation  
\cite{Bose:2008ew,DeSantiago:2011qb,Sahni:2015hbf,Mishra:2018tki}. 
Additionally, note that in the inflation realization in the context of  
Galileon and Horndeski theories, the role of the non-canonical kinetic term is 
also crucial 
\cite{Kobayashi:2010cm,Burrage:2010cu,Ohashi:2012wf,Tsujikawa:2014mba, Starobinsky:2016kua, 
Sebastiani:2017cey, Koutsoumbas:2017fxp}. The form of the non-canonical terms 
can vary significantly, since there are many plausible models, including 
different ways to achieve compactification. The recent cosmological data, 
however, together with
the requirement to avoid fine-tuning and non-physical solutions, severely
constrain the available possibilities.

On the other hand, an alternative way to improve the inflationary observables 
is by introducing an extra parameter as an exponent in the known potential 
forms, and thus affecting their steepness. In this way the dynamics of the 
scalar field can be additionally deformed, offering an alternative way to bring 
the tensor-to-scalar ratio to lower values without ruining the necessary 
spectral index 
\cite{Geng:2015fla,Rezazadeh:2015dia,Geng:2017mic,Ahmad:2017itq,Agarwal:2017wxo,
Skugoreva:2019blk,Das:2019ixt,Lima:2019yyv,Nojiri:2017ncd,Bhattacharya:2018xlw,Banerjee:2018kcz}.
 
One possible disadvantage of the above inflationary models, namely
those with 
non-canonical terms and those with extra steepness parameter in the potential, 
is 
that the parameter values needed for acceptable observables   are  
quite unnatural (with ``natural'' meaning the standard values of canonical 
kinetic term, no  extra steepness parameter, and sub-Planckian potential 
values) and 
hard to be justified from the field-theoretical point of view. In particular, 
the non-canonical exponents need to be large, or the mass and 
potential parameters take trans-Planckian values. Hence, in this 
work, we are interested in  studying a combination of the above models, 
specifically 
introducing   a scalar field with non-canonical
kinetic terms on top of a deformed-steepness potential with an extra parameter. 
As we will show, this  
enhances the range of solutions and leads to very satisfactory observables, for  sets of model parameters significantly closer to the natural ones,
that we proceed to identify and classify.

\section{Non-canonical inflation with deformed-steepness potentials}

In this section we present the scenario of non-canonical inflation with 
deformed-steepness potentials. We will focus on the usual non-canonical 
Lagrangian,  which is well justified theoretically, and takes the form  
\cite{ArmendarizPicon:1999rj,Garriga:1999vw,Mukhanov:2005bu,
Unnikrishnan:2008ki,Li:2012vta,Unnikrishnan:2012zu},
\be
\label{ourLagr}
  \mathcal{L}(\phi, X)=X \left(\frac{X}{M^4} \right)^{\alpha
    -1}-V(\phi),
 \ee
 where $X=\frac{1}{2} \partial_{\mu} \phi \partial^{\mu} \phi$ is the kinetic energy 
of the scalar field, and thus the action of the scenario reads
\be \label{QNCKT action}
\mathcal{S}=\int d^{4}x \sqrt{-g} \left[ M^{2}_{pl} \frac{R}{2} +X 
\left(\frac{X}{M^4} \right)^{\alpha -1}-V(\phi) \right].
\ee
The parameter $M$ has dimensions of mass and determines the scale in which the 
non-canonical effects become significant, while    $M_{pl}$ is the Planck 
mass.
Concerning the potential, in this work we will consider the deformed-steepness 
potential that was introduced in 
\cite{Geng:2015fla,Geng:2017mic}, namely
\be
\label{poten11}
V(\phi)=V_{0}\,e^{-\lambda \phi^{n}/M^{n}_{pl}},
\ee
with $V_0$ and $\lambda$ the usual potential parameters and $n$ the new 
exponent 
parameter that determines the deformed-steepness.

We consider a homogeneous and isotropic  
 flat   Friedmann-Robertson-Walker (FRW) 
  metric  of the form
\begin{equation}
\label{FRWmetric}
ds^{2}=-dt^{2}+a^{2}(t)\delta_{ij}dx^{i}dx^{j}\,,
\end{equation}
where $a(t)$ is the scale factor. 
Variation of the action   (\ref{QNCKT action}) in terms of the metric 
gives
the following Friedmann equations 
\be
\label{Fried1}
H^2=\frac{1}{3 M^{2}_{pl}} \left[(2\alpha -1)X \left(\frac{X}{M^4} 
\right)^{\alpha-1}+V_{0}\,e^{-\lambda \phi^{n}/M^{n}_{pl}} \right] 
\ee
\be 
\label{Fried2}
\dot H=- \frac{1}{M^{2}_{pl}} \alpha X\left(\frac{X}{M^4} \right)^{\alpha-1},
\ee
 where  $H=\frac{\dot a}{a}$ is the  Hubble parameter. Additionally, variation 
in terms of the scalar field  leads to the Klein-Gordon equation
\be
\ddot \phi +\frac{3H\dot \phi}{2\alpha -1}-\frac{\lambda n 
\phi^{n-1}V_{0}\,e^{-\lambda  
\phi^{n}/M^{n}_{pl}}}{\alpha (2\alpha -1)M^{n}_{pl}} 
\left(\frac{2M^4}{\dot{\phi ^2}} \right)^{\alpha -1}=0.
\label{KLGord}
\ee
Note that one can write the above equation in the form of the usual 
conservation equation $\dot{\rho}_{\phi}+3H(\rho_\phi+p_\phi)$, using the 
definitions 
\begin{eqnarray}
&&
\rho_{\phi} = (2\alpha -1)X \left(\frac{X}{M^4} 
\right)^{\alpha-1}+V_{0}\,e^{-\lambda \phi^{n}/M^{n}_{pl}}
\nonumber\\ \nonumber\\
&& 
p_{\phi} = X \left(\frac{X}{M^4} \right)^{\alpha-1}-V_{0}\,\,e^{-\lambda 
\phi^{n}/M^{n}_{pl}}.
\end{eqnarray}

In every inflationary scenario the important quantities are the 
inflation-related observables, namely the scalar spectral index of the 
curvature 
perturbations $n_\mathrm{s}$ and its running $\alpha_\mathrm{s} \equiv d 
n_\mathrm{s}/d 
\ln k$, with $k$   the measure of the wave number $\Vec{k}$,
the tensor spectral 
index $n_\mathrm{T}$ and its running, as well as the tensor-to-scalar ratio 
$r$. In a given scenario  these quantities depend on the model parameters, and 
hence  confrontation with observational data can lead to constraints on these  
model parameters. 

In order to extract the relations for the inflation-related observables,
  a detailed and thorough perturbation analysis is needed. In the simple case 
of canonical fields minimally coupled to gravity, and introducing the slow-roll 
parameters, full perturbation analysis indicates that the  
inflationary observables can be expressed solely in terms of the scalar potential and 
its derivatives  \cite{Lidsey:1995np, Lyth:1998xn,Martin:2013tda}. 
However, in the case where non-canonical terms or forms of non-minimally 
coupling are present, as well as in the case where the potential itself is 
absent (as for instance in modified gravity inflation),
  one should instead introduce the Hubble slow-roll parameters 
$\epsilon_n$ 
(with $n$ positive integer),  defined as 
\cite{Kaiser:1994vs,Sasaki:1995aw,Martin:2013tda,Woodard:2014jba}
\begin{eqnarray}
\epsilon_{n+1}\equiv \frac{d\ln |\epsilon_n|}{dN},
\label{epsilonnn}
\end{eqnarray}
where $N\equiv\ln(a/a_{ini})$ is the 
e-folding number, and 
 $\epsilon_0\equiv H_{ini}/H$, where
  $a_{ini}$ is the initial scale factor  with 
$H_{ini}$ the corresponding Hubble parameter  (as usual inflation ends when 
$\epsilon_1=1$). Thus, the 
first three $\epsilon_n$ are found to  be  
\begin{eqnarray}
\label{e1}
&&\!\!\!\!\!\!\!\!\!\!\!\!\!\!
\epsilon_1\equiv-\frac{\dot{H}}{H^2}, 
\\
&&\!\!\!\!\!\!\!\!\!\!\!\!\!\!
\epsilon_2 \equiv  \frac{\ddot{H}}{H\dot{H}}-\frac{2\dot{H}}{H^2},
\label{e2}\\
&&\!\!\!\!\!\!\!\!\!\!\!\!\!\!
\epsilon_3 \equiv
\left(\ddot{H}H-2\dot{H}^2\right)^{-1} 
 \left[\frac{H\dot{H}\dddot{H}-\ddot{H}(\dot{H}
^2+H\ddot{H}) } { H\dot { H } }-\frac{2\dot{H}}{H^2}(H\ddot{H}-2\dot{H}^2)
\right].
\label{e3}
\end{eqnarray}
With these definitions, the basic inflationary observables are given as
  \cite{Martin:2013tda} 
\begin{eqnarray}
r &\approx&16 c_{s} \epsilon_1 ,
\label{r}\\
 n_\mathrm{s} &\approx& 1-2\epsilon_1 - \epsilon_2  ,
\label{ns} \\
\alpha_\mathrm{s} &\approx & -2 \epsilon_1 \epsilon_2 - \epsilon_2 \epsilon_3  ,
\label{as}\\
n_\mathrm{T} &\approx& -2\epsilon_1  ,
\label{nT}
\end{eqnarray}

where the sound speed is defined as \cite{Unnikrishnan:2012zu}
\be \label{spofsoun}
c^{2}_{s}=\left [\frac{(\partial \mathcal{L}/\partial X)}{(\partial 
\mathcal{L}/\partial X)+(2X)(\partial^{2} \mathcal{L}/\partial X^{2})} \right ],
\ee
which for the 
 Lagrangian   (\ref{ourLagr}) gives
\be 
c^{2}_{s}=\frac{1}{2\alpha -1}.
\ee

In the scenario of non-canonical inflation with deformed-steepness potentials, 
described by equations (\ref{Fried1})-(\ref{KLGord}),
 the dynamics, i.e. the Hubble function,  is determined by the 
 parameters $\alpha$ and $M$ related to ``non-canonicality'', by the standard 
potential parameters $V_0$ and $\lambda$, alongside the  
deformed-steepness parameter $n$. Hence, we deduce that the above inflationary 
observables (\ref{r})-(\ref{nT}) will be determined by these model parameters 
too. Nevertheless, we should mention that these parameters are not 
independent 
since they are related through the observed value of the amplitude of 
scalar perturbation $\mathcal{A}_{S}$. In particular, for the action 
(\ref{QNCKT action}) one finds \cite{Unnikrishnan:2012zu}
\be
\mathcal{P}_{S} = \left (\frac{1}{72\pi^{2}c_{_s}}\right )\left 
\{
 M_{pl}^{-2(5\alpha-6)} 
 \left[\frac{\alpha 6^{\alpha}}{M^{4(\alpha-1)}}\right ]
 \left 
[\frac{V(\phi)^{5\alpha - 2}}{V'(\phi)^{2\alpha}}\right ]
\right \}^{\frac{1}{2\alpha - 1}},
\ee
which for our potential (\ref{poten11})
leads to
 \be 
\mathcal{P}_{S}=\mathcal{A}_{S}\left [ \,e^{-(3\alpha -2)\lambda 
\frac{\phi^{n}}{M_{pl}^{n}}}\left(\frac{\phi}{M_{pl}}\right )^{2\alpha 
(1-n)}\right]^{\frac{1}{2\alpha - 1}},
\ee
with
\be
\mathcal{A}_{S} = \left (\frac{1}{72\pi^{2}c_{_s}}\right )\left 
\{
 M_{pl}^{4(1-2\alpha)} \left[\frac{\alpha 6^{\alpha}}{M^{4(\alpha-1)}}\right ]
 \left 
[\frac{V_{0}^{3\alpha - 2}}{(\lambda n)^{2\alpha}}\right ]
\right \}^{\frac{1}{2\alpha - 1}}.
\ee
Since  we know that Planck observations give \cite{Akrami:2018odb}
\be
\mathcal{A}_{Sobs}\simeq 2.09052 \cdot 10^{-9},
\label{obscon}
\ee
in the following analysis we set  $V_0$ at will while $\lambda$ arises 
from the satisfaction of the above observational constraint.

In the next section we investigate in detail the effect of each parameter on 
the inflationary observables, and we will show which combinations can bring the 
predictions deep inside the observational contours.

\section{Results}

In this section, we investigate the inflationary observables in the scenario of 
non-canonical inflation with deformed-steepness potentials. In particular, we 
desire to see how the scalar spectral index  $n_\mathrm{s}$ and the 
tensor-to-scalar ratio $r$ are affected by the model parameters. Since the 
involved equations  (\ref{Fried1})-(\ref{KLGord}), the slow-roll parameters  
(\ref{e1})-(\ref{e3}) and the observables expressions
(\ref{r})-(\ref{nT}) are in general too complicated to admit analytical 
solutions, we investigate them numerically.

Specifically, for a given set of parameter values we impose the conditions 
for  $\phi$, $\dot \phi$ and $H$  corresponding    
to  $\epsilon_i\ll1$. According to (\ref{e1})-(\ref{e3}), such 
suitably ``small'' $\epsilon_i$  imply $H\sim const.$ and thus almost 
exponential expansion, which is the typical feature of inflation. 
We 
evolve the system and we determine the end of inflation by demanding at 
least one of the
$\epsilon_i$ to become 1 (cases of eternal inflation, in which all $\epsilon_i$ 
remain smaller than 1  are considered non-physical), and 
thus by imposing the desired e-folding number $N$ we extract the time at the 
beginning of inflation. Hence, we can use the corresponding Hubble parameter 
to 
calculate the inflationary observables corresponding to the given parameter 
values and the imposed  e-folding number $N$.

\begin{figure}[ht]
\begin{center}
\hskip-1cm
\includegraphics[scale=.59]{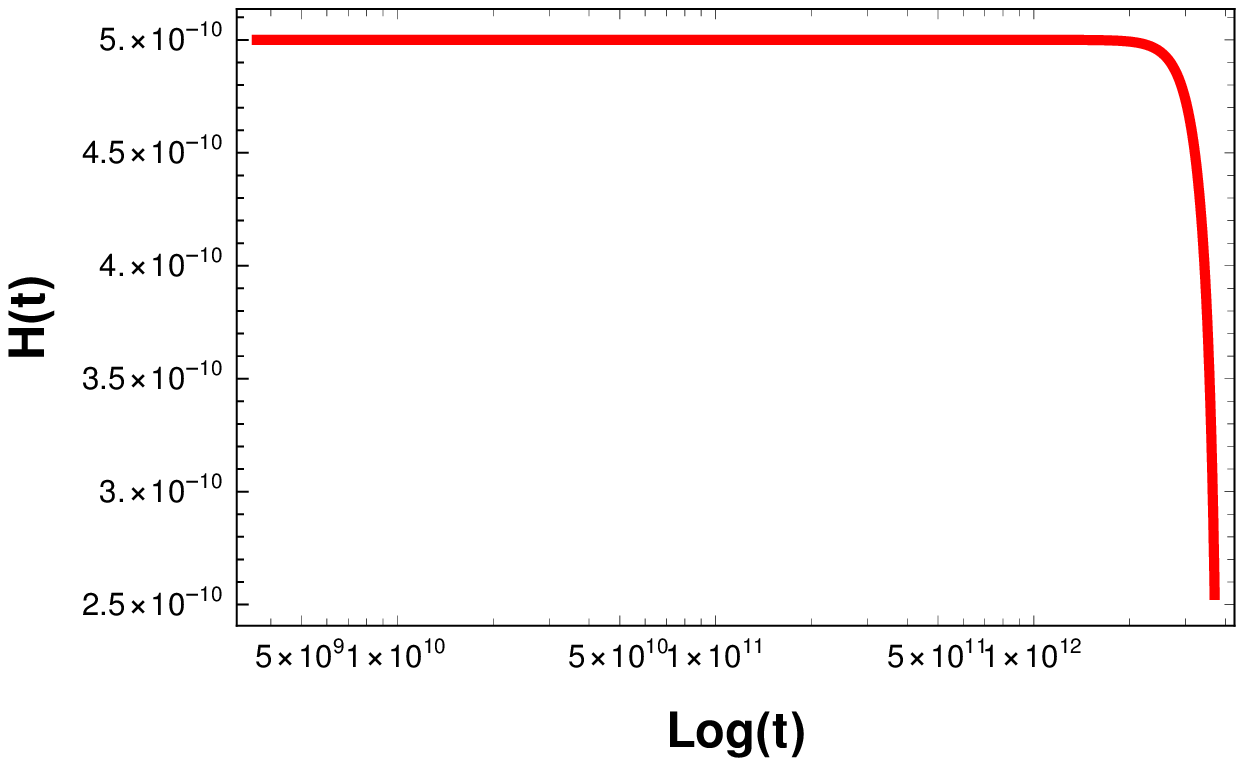}
\includegraphics[scale=.55]{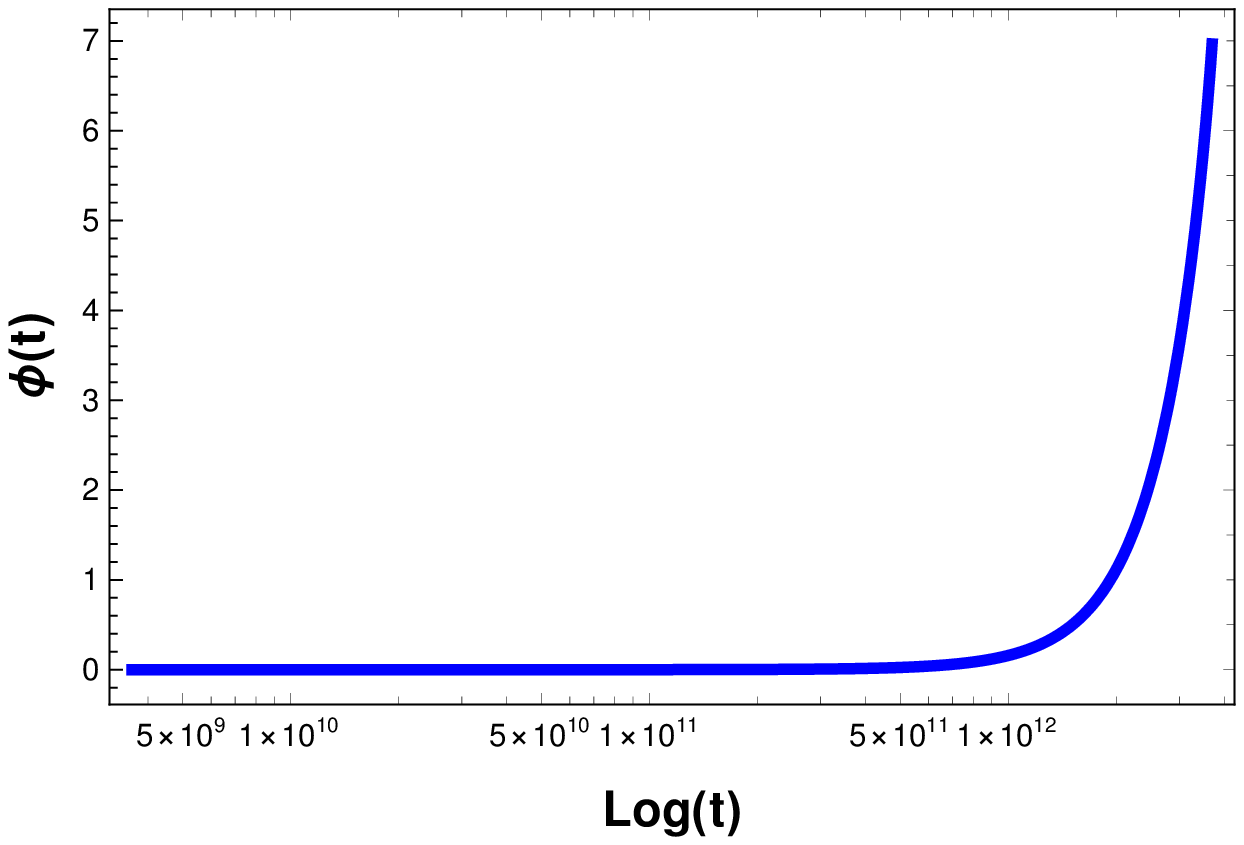}\\
\includegraphics[scale=.55]{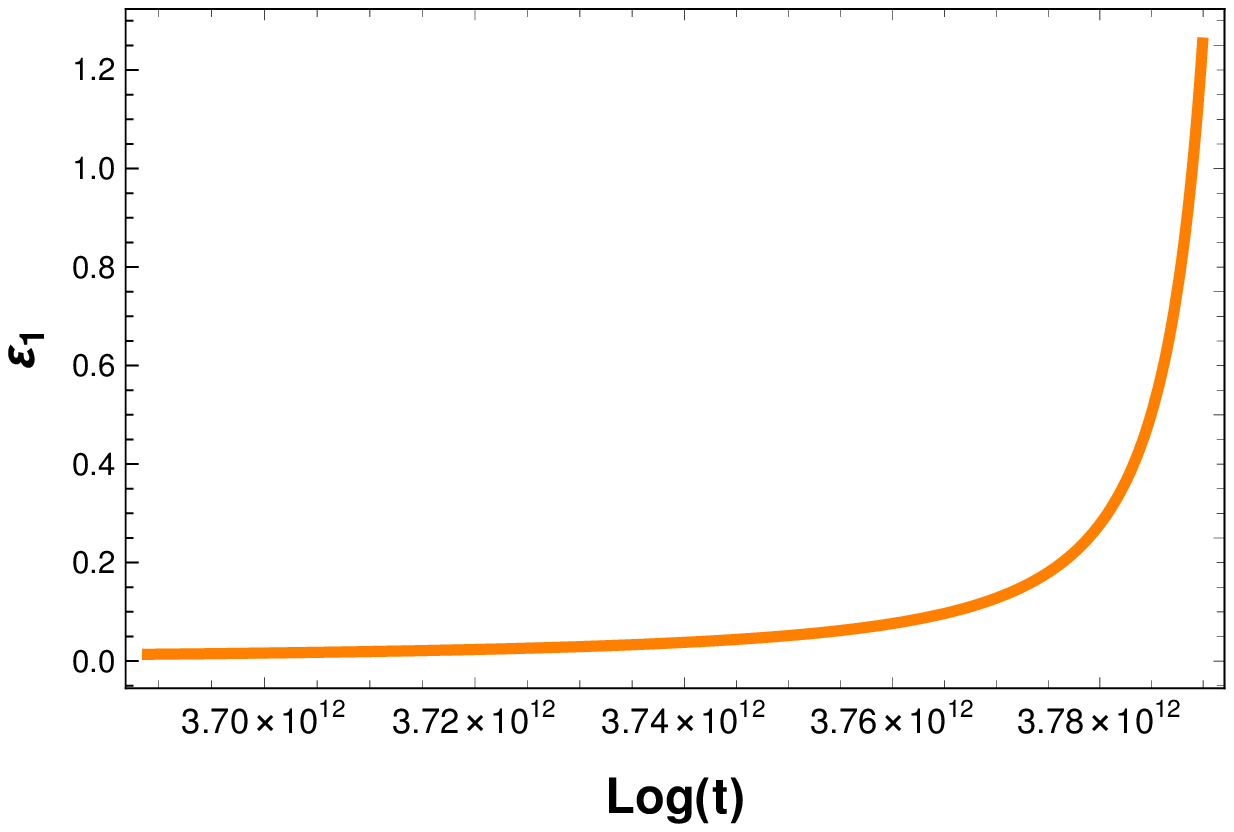}
\includegraphics[scale=.55]{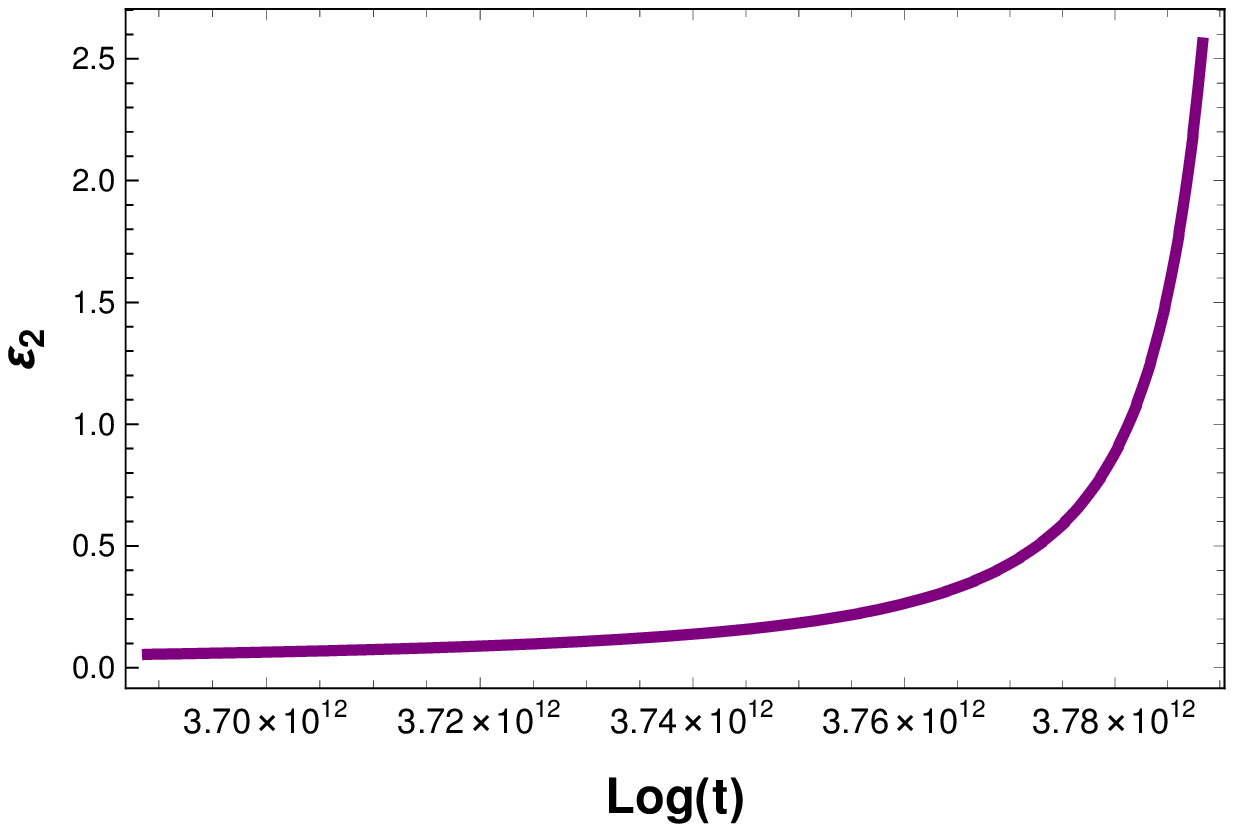}
\end{center}
\caption{{\it{ Time evolution
 of the Hubble parameter, of the scalar field $\phi$, and of the Hubble 
slow-roll parameters $\epsilon_{1}$ and $\epsilon_{2}$ in the case where $\alpha 
=2$ and $n=3$. Initially, and for a suitable time interval, $\epsilon_i\ll1$ 
and thus according to (\ref{e1})-(\ref{e3}) $H$ is almost constant, which   
corresponds to 
inflation realization,  while  as time passes  $\epsilon_i$ increase and when 
one of them reaches 1 exponential expansion is not the case anymore ($H$ 
decreases) and 
inflation ends.
 }}}
\label{fige}
\end{figure}

In order to show this in a more transparent way, in Fig. \ref{fige} we depict 
the evolution of $H$, $\phi$,  $\epsilon_{1}$ and $ \epsilon_{2}$ in terms of 
time. As we observe, initially and for a suitable time interval, $\epsilon_i$ 
are small and thus $H$ is almost constant, which indeed corresponds to 
inflation realization,  while  as time passes, $\epsilon_i$ increase and when 
one of them reaches 1 exponential expansion is not the case anymore ($H$ 
decreases) and 
inflation ends.  Hence, the imposed initial conditions have been 
chosen in      order for a successful inflation to occur, which requires 
two conditions: i) the initial $\epsilon_i$ to be $\epsilon_i\ll1$, and 
ii) before one of them reaches 1 and inflation ends to have obtained the 
required ``suitable time interval'', namely expansion of 50-70 e-foldings. We 
mention that such background solutions are quite common for the studied range of 
the model parameters.

\begin{figure}[ht]
\begin{center}
\hskip-1cm
\includegraphics[scale=.41]{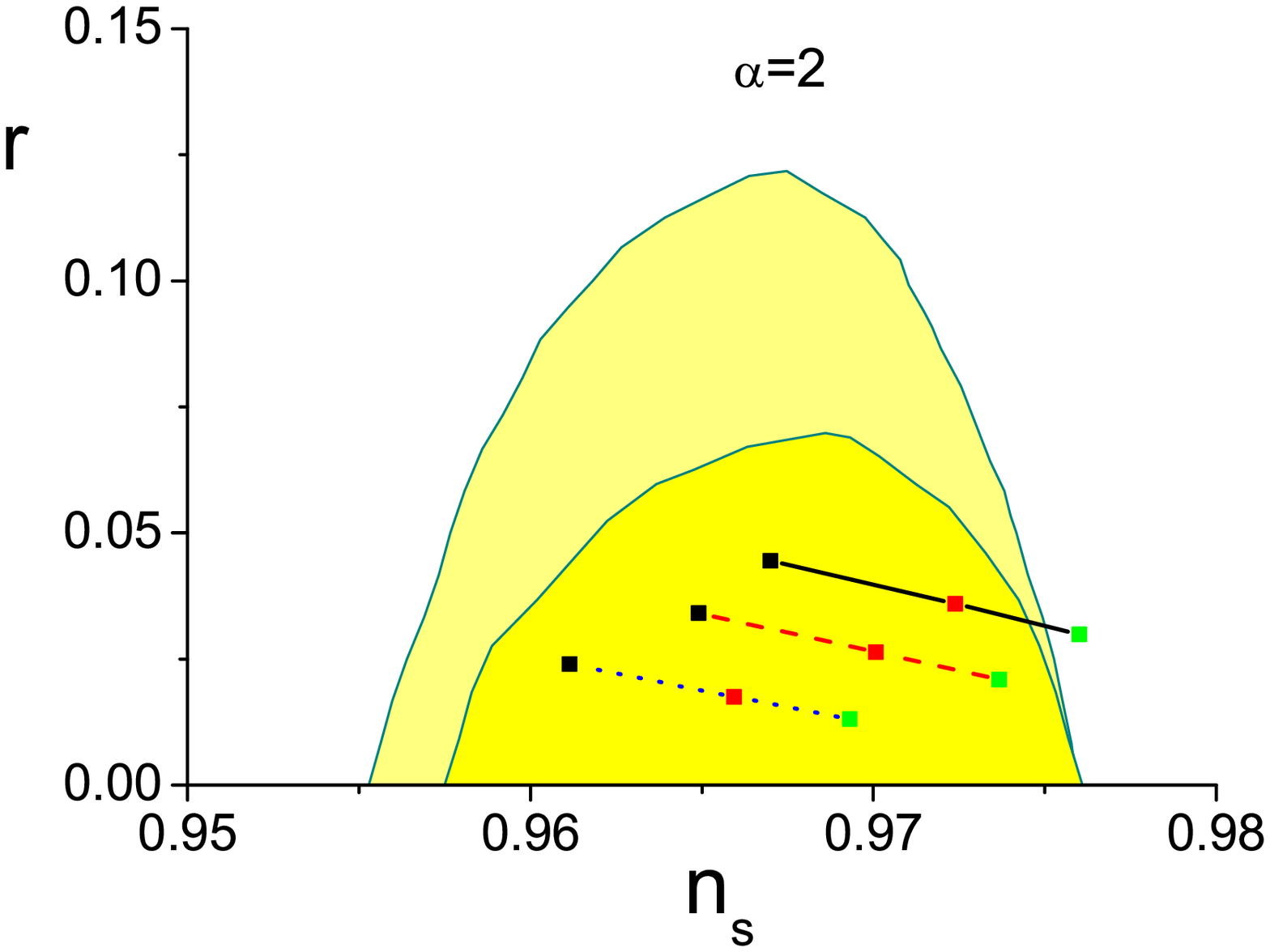}
\includegraphics[scale=.41]{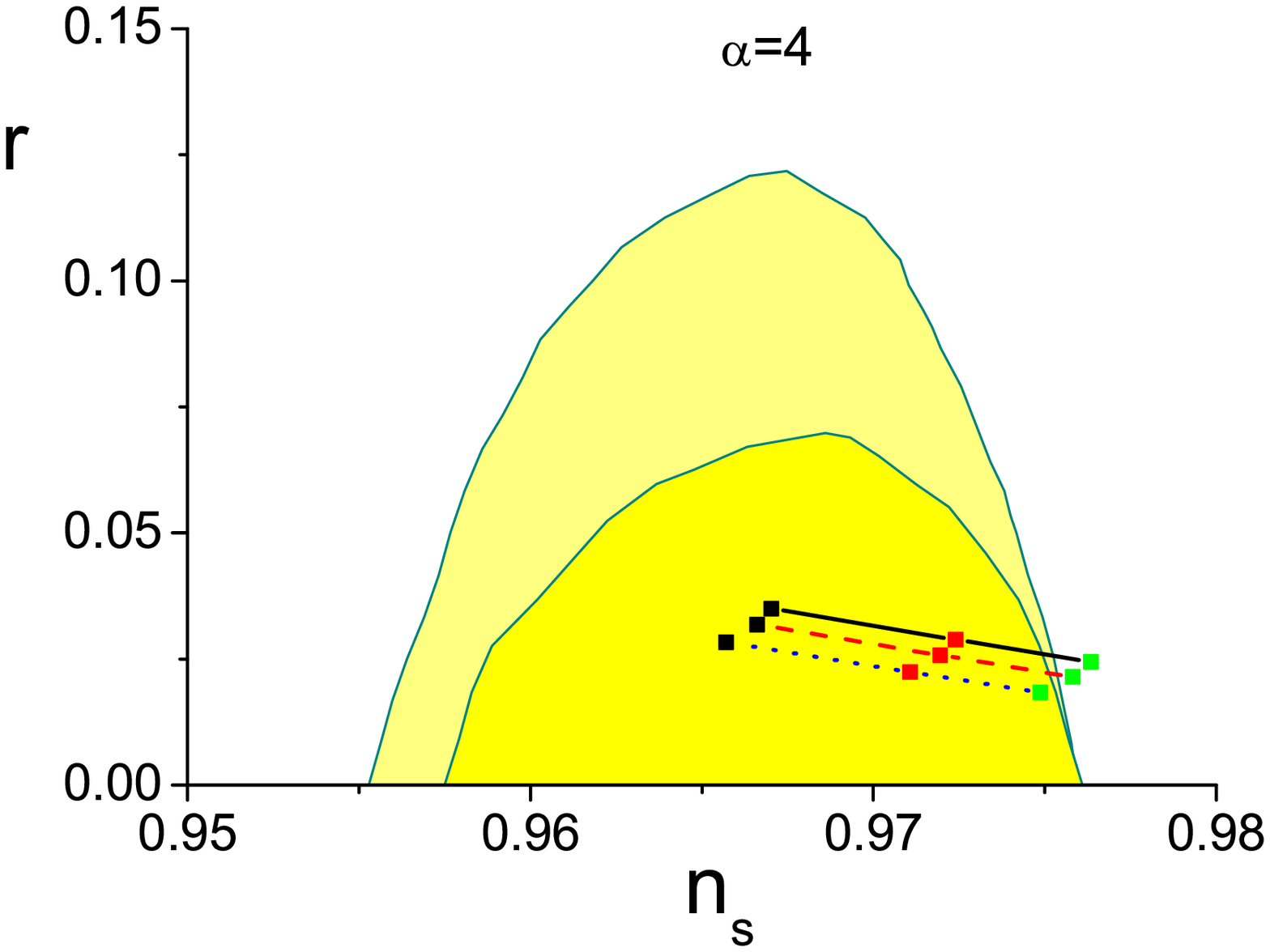}
\includegraphics[scale=.41]{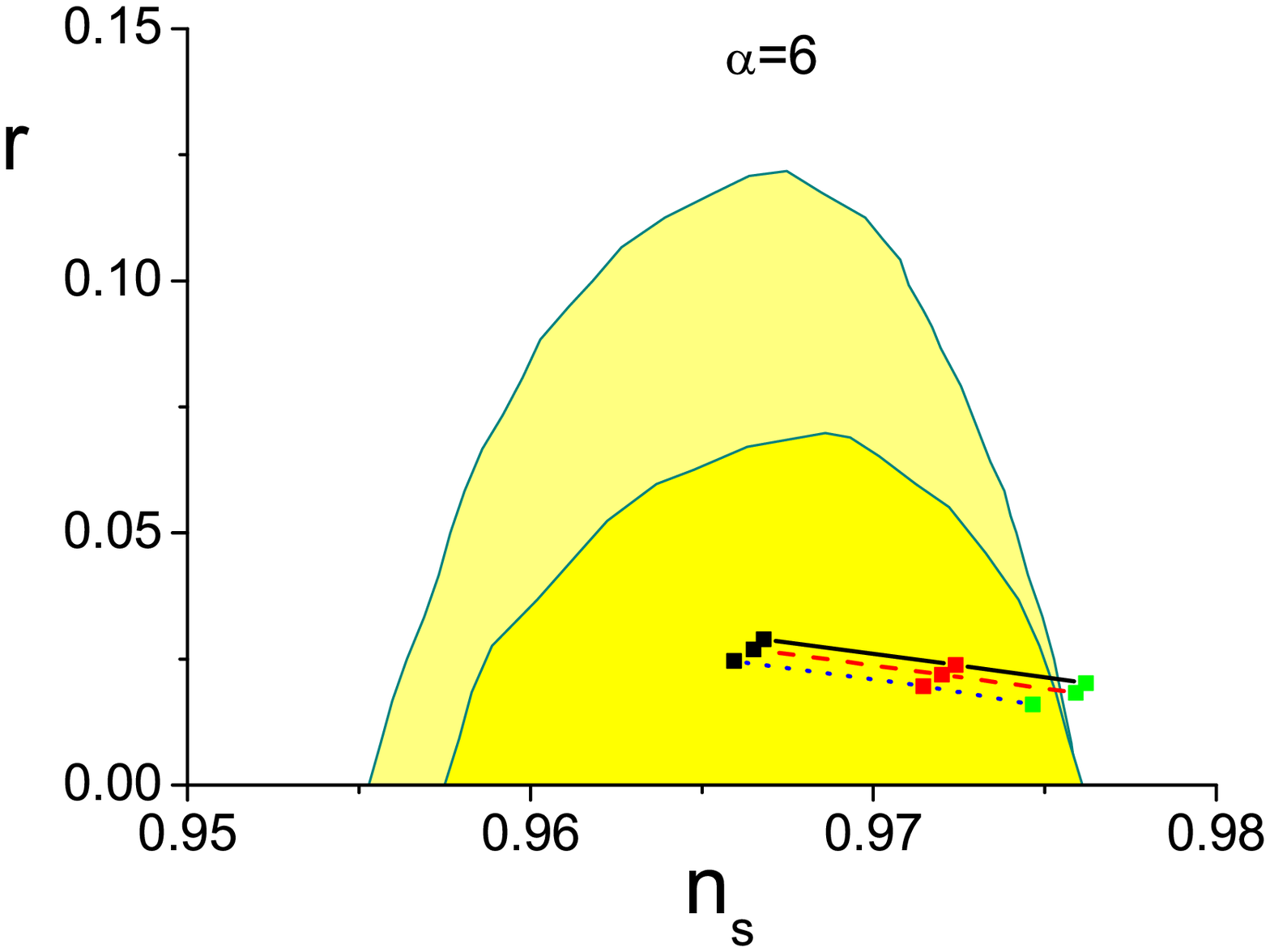}
\end{center}
\caption{
{\it{  1$\sigma$ (yellow) and 2$\sigma$ (light yellow) contours for Planck 2018
results (Planck $+TT+lowP$)  \cite{Akrami:2018odb}, on the
$n_{\mathrm{s}}-r$ plane.
Furthermore, we 
depict  the predictions of Table \ref{Tablebasic}, of the scenario at hand for 
various values of the   
the non-canonical 
parameter $\alpha$ and the deformed-steepness parameter $n$,  keeping
fixed $M=10^{-6} M_{pl}$ and $V_{0}=  10^{-16} M_{pl}^4$, with $M_{pl}=10^{18}$ 
GeV,  determining  the value of $\lambda$   
 through the observational constraint (\ref{obscon}). In all graphs,  the 
black-solid line corresponds to $n=3$, the red-dashed line to $n=4$ and the 
blue-dotted line to $n=5$. In every line 
the first (black) point 
corresponds to  e-folding number $N=50$, the middle (red) point to $N=60$, and 
the third (green) to $N=70$.  
}}}
\label{figalphan}
\end{figure}

We start our investigation by examining the effect of the non-canonical 
parameter $\alpha$ and the deformed-steepness parameter $n$. Therefore,  we
fix $M$ and $V_0$ at theoretically motivated values  and we calculate  
$n_\mathrm{s}$ and  $r$ for various combinations of 
 $\alpha$ and $n$, adjusting suitably only the value of $\lambda$ in 
order to satisfy   the   observational constraint (\ref{obscon}), and for  the
e-folding number $N$ taking, as usual, the values 50, 60 and 70. 
In Table \ref{Tablebasic}  we summarise the obtained observable predictions.
Additionally, in order to present the information in a more transparent way 
that 
allows comparison with observational data, in Fig. \ref{figalphan} we depict 
the results  of Table \ref{Tablebasic} on top of the 1$\sigma$ and 
2$\sigma$ contours of the Planck 2018 data \cite{Akrami:2018odb}.

\begin{table}
\vskip0.1cm
\hskip-1.cm
\begin{tabular}{|l|l|l|l|}
\hline
\multicolumn{4}{|l|}{$\alpha=2$,\ $n=3$,\ $\lambda=7.54\cdot 
10^{-6}$} \\ 
\hline
 $N$    &  50   &  60   &  70   \\ \hline
 $r$    &   0.0445   &   0.0359   &    0.0299   \\ \hline
  $n_{s}$    &  0.9670  &  0.9723   &  0.9761   \\ \hline
\end{tabular}
\hskip0.1cm
\begin{tabular}{|l|l|l|l|}
\hline
\multicolumn{4}{|l|}{$\alpha=2$,\ $n=4$,\ $\lambda=5.65\cdot 
10^{-6}$} \\ 
\hline
 $N$    &  50   &  60   &  70   \\ \hline
 $r$    &   0.0341   &   0.0263   &    0.0209  \\ \hline
  $n_{s}$    &  0.9649  &  0.9700   &  0.9736  \\ \hline
\end{tabular}
\hskip0.1cm
\begin{tabular}{|l|l|l|l|}
\hline
\multicolumn{4}{|l|}{$\alpha=2$,\ $n=5$,\ $\lambda=4.52\cdot 
10^{-6}$} \\ 
\hline
 $N$    &  50   &  60   &  70   \\ \hline
 $r$    &   0.0240   &   0.0174   &    0.0130   \\ \hline
 $n_{s}$    &  0.9611  &  0.9659   &  0.9693   \\ \hline
\end{tabular}
\vskip0.1cm
\hskip-1.cm
\begin{tabular}{|l|l|l|l|}
\hline
\multicolumn{4}{|l|}{$\alpha=4$,\ $n=3$,\ $\lambda=7.61\cdot 
10^{-6}$} \\ 
\hline
 $N$    &  50   &  60   &  70   \\ \hline
 $r$    &   0.0350   &   0.0288   &    0.0244 \\ \hline
 $n_{s}$    &  0.9670  &  0.9724   &  0.9763 \\ \hline
\end{tabular}
\hskip0.1cm
\begin{tabular}{|l|l|l|l|}
\hline
\multicolumn{4}{|l|}{$\alpha=4$,\ $n=4$,\ $\lambda=5.71\cdot 
10^{-6}$} \\ 
\hline
 $N$    &  50   &  60   &  70   \\ \hline
 $r$    &   0.0318   &   0.0257   &    0.0214   \\ \hline
 $n_{s}$    &  0.9666 &  0.9719   &  0.9758  \\ \hline
\end{tabular}
\hskip0.1cm
\begin{tabular}{|l|l|l|l|}
\hline
\multicolumn{4}{|l|}{$\alpha=4$,\ $n=5$,\ $\lambda=4.57\cdot 
10^{-6}$} \\ 
\hline
 $N$    &  50   &  60   &  70   \\ \hline
 $r$    &   0.0282   &   0.0224   &    0.0183   \\ \hline
 $n_{s}$    &  0.9657 &  0.9710   &  0.9748  \\ \hline
\end{tabular}
\vskip0.1cm
\hskip-1.cm
\begin{tabular}{|l|l|l|l|}
\hline 
\multicolumn{4}{|l|}{$\alpha=6$,\ $n=3$,\ $\lambda=8.25\cdot 
10^{-6}$} \\ 
\hline 
 $N$    &  50   &  60   &  70   \\ \hline 
 $r$    &   0.0289   &   0.0238   &    0.0202   \\ \hline 
 $n_{s}$    &  0.9668  &  0.9724   &  0.9762   \\ \hline 
\end{tabular}
\hskip0.1cm
\begin{tabular}{|l|l|l|l|}
\hline
\multicolumn{4}{|l|}{$\alpha=6$,\ $n=4$,\ $\lambda=6.19\cdot 
10^{-6}$} \\ 
\hline
 $N$    &  50   &  60   &  70   \\ \hline
 $r$    &   0.0268   &   0.0218   &    0.0183   \\ \hline
 $n_{s}$    &  0.9660  &  0.9700   &  0.9759   \\ \hline
\end{tabular}
\hskip0.1cm
\begin{tabular}{|l|l|l|l|}
\hline
\multicolumn{4}{|l|}{$\alpha=6$,\ $n=5$,\ $\lambda=4.95\cdot 10^{-6}$} \\ 
\hline
 $N$    &  50   &  60   &  70   \\ \hline
 $r$    &   0.0245   &   0.0197   &    0.0163   \\ \hline
 $n_{s}$    &  0.9653  &  0.9714   &  0.9746  \\ \hline
\end{tabular}
\caption{Predictions for the   scalar spectral index  $n_\mathrm{s}$ and 
the tensor-to-scalar ratio $r$ for the scenario of non-canonical inflation with 
deformed-steepness potential, for various combinations of $\alpha$ and $n$, 
adjusting the values of $\lambda$ in 
order to satisfy   the   observational constraint (\ref{obscon}), and for 
e-folding number $N$ equal to 50, 60 
and 70. For this Table we fix $M=10^{-6} M_{pl}$ and $V_{0}=10^{-16}  M_{pl}^4$ 
 , with $M_{pl}=10^{18}$ $GeV$, while the value of $\lambda$ is 
determined through the observational constraint (\ref{obscon}).} 
\label{Tablebasic} 
\end{table}

A general observation is that the predictions of the scenario at hand lie well 
inside the  1$\sigma$ region of the Planck 2018 data, without the need to use 
large values for the non-canonical parameter $\alpha$ or the  
deformed-steepness 
parameter $n$, which was indeed the main motivation behind the present work.  
Additionally, the predictions of the scenario are better, compared to the 
simple non-canonical models, as well as to the simple  deformed-steepness 
models.

Concerning the specific features, we find the following:
For any given set of model-parameters, increasing the e-folding values $N$ leads 
to increased $n_{\mathrm{s}}$ and decreased $r$, as is usual in the majority 
of inflationary scenarios. Now, for a given $\alpha$, as $n$ increases both 
$n_{\mathrm{s}}$ and   $r$ also increase. On the other hand, for a given 
$n$, as $\alpha$ increases there is no particular tendency for 
$n_{\mathrm{s}}$ and   $r$. However, for larger $\alpha$, such as $\alpha=6$,
the 
effect of $n$ is less significant and the different curves actually coincide. 
Moreover, as $\alpha$ grows,    $\lambda$ slightly increases, while as $n$ grows
 $\lambda$ slightly decreases.

\begin{figure}[!h]
\begin{center}
\hskip-1cm
\includegraphics[scale=.41]{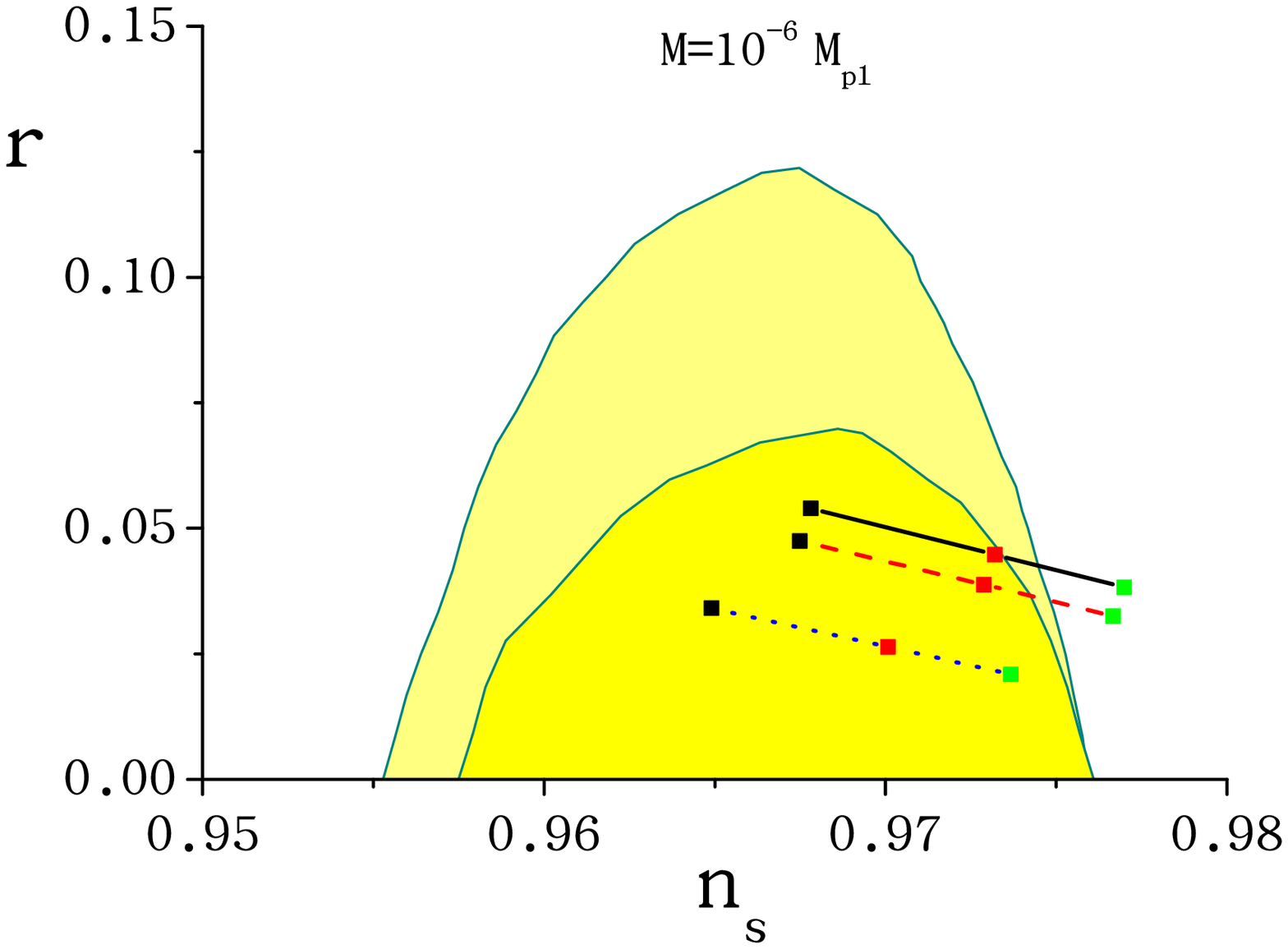}
\includegraphics[scale=.41]{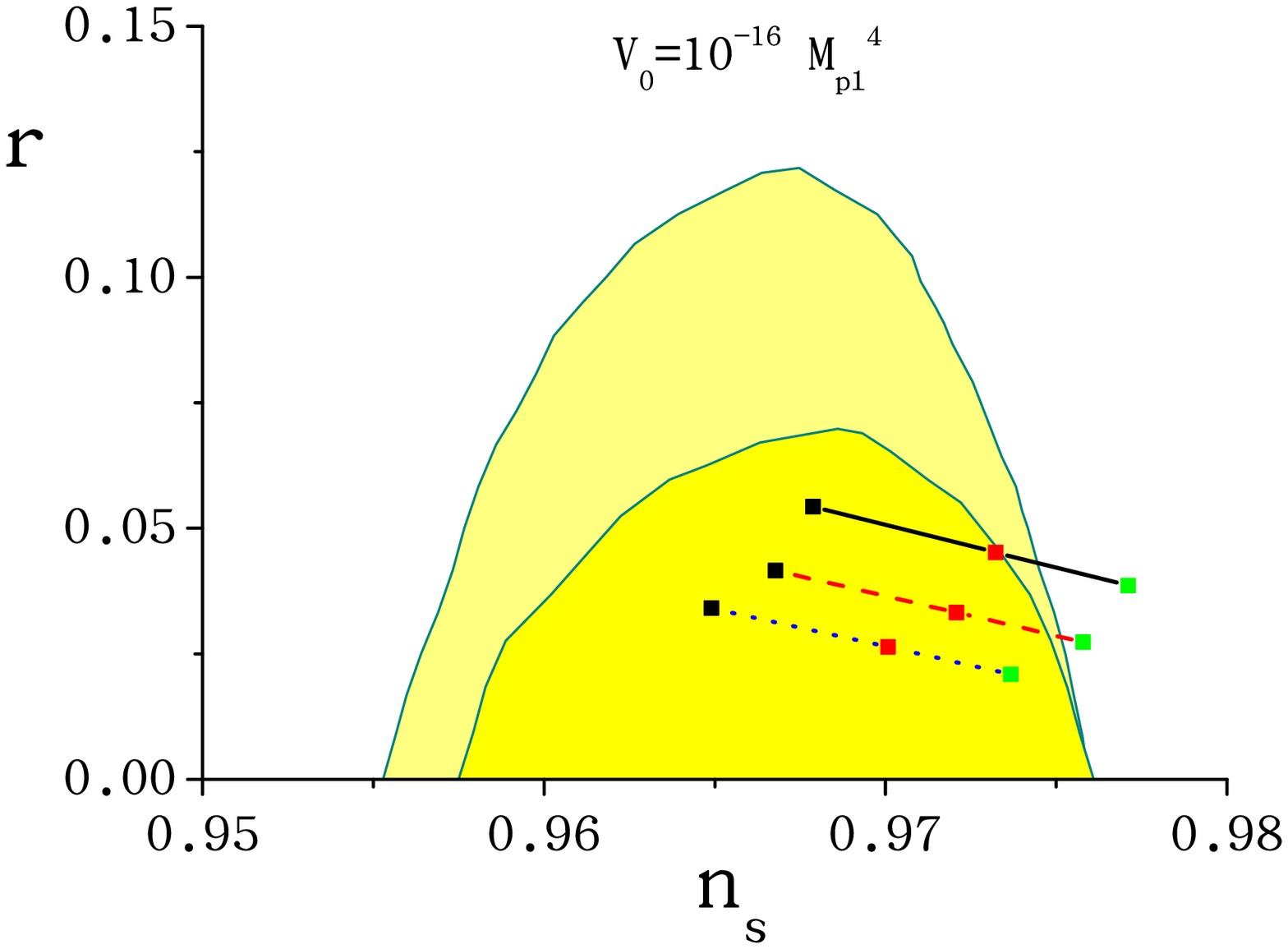}
\end{center}
\caption{
{\it{  1$\sigma$ (yellow) and 2$\sigma$ (light yellow) contours for Planck 2018
results (Planck $+TT+lowP$)  \cite{Akrami:2018odb}, on the
$n_{\mathrm{s}}-r$ plane.
Moreover, we 
depict  the predictions of   Table \ref{Table2}, of the 
scenario at hand   
for fixed $\alpha=2$, $n=4$.  In every line the first (black) point 
corresponds to  e-folding number $N=50$, the middle (red) point to $N=60$, and 
the third (green) to $N=70$.
{\bf{Left panel}}: 
Fixed  $M=10^{-6} M_{pl}$.  Black-solid for 
$V_{0}=10^{-18} M_{pl}^4$, red-dashed for $V_{0}=10^{-17} M_{pl}^4$, blue-dotted 
for $V_{0}=10^{-16} M_{pl}^4$.
{\bf{Right panel}}:
Fixed $V_{0}=10^{-16} M_{pl}^4$.  Black-solid for 
$M=10^{-7} M_{pl}$, red-dashed for  $M=5\cdot 10^{-7} M_{pl}$, red-dotted 
for  $M=10^{-6} M_{pl}$.
In all cases    the 
value of $\lambda$   is determined  through the observational constraint 
(\ref{obscon}), while $M_{pl}=10^{18}$ $GeV$.
}}}
\label{fig2}
\end{figure}

 We proceed by investigating the effect on the observables of the parameters $M$ 
and $V_0$, which determine the scale of non-canonicality and of the potential, 
respectively, keeping in mind that
the scale of inflation in theoretically
motivated constructions can be anywhere from just below the
unification scale (mostly Grand Unification), to energies as low as
the scales within reach of the LHC (see e.g. \cite{German:2000yz}), with many
possibilities in between, that can be linked i.e. to different stages
of symmetry breaking.
Without loss of generality we fix $\alpha=2$ and 
$n=4$   and we calculate  
$n_\mathrm{s}$ and  $r$ for 
e-folding number $N$ being as usual 50, 60 and 70. We first additionally 
fix 
$M$ 
and change $V_0$, and then we fix $V_0$ and change $M$. In all cases 
we adjust  the value of $\lambda$ in 
order to satisfy   the   observational constraint (\ref{obscon}).
We summarise the 
obtained observable predictions 
in Table \ref{Table2}.  
Moreover, in order to present the results in a more transparent way, in 
Fig. \ref{fig2} we depict 
the results  of Table \ref{Table2} on top of the 
  1$\sigma$ and 
2$\sigma$ contours of the Planck 2018 data \cite{Akrami:2018odb}.

\begin{table}[ht]
\begin{tabular}{|l|l|l|l|l|l|l|l|l|l|}
\hline
\multicolumn{10}{|l|}{$\alpha=2$, \ \ \ \ $n=4$, \ \ \ \ $M=10^{-6}M_{pl}$} \\ \hline
 $V_{0}$ &  \multicolumn{3}{l|}{$  10^{-16} M_{pl}^4$ } & 
\multicolumn{3}{l|}{$  10^{-17} M_{pl}^4$  } & 
\multicolumn{3}{l|}{$  10^{-18} M_{pl}^4$} \\ \hline
$\lambda$ &  \multicolumn{3}{l|}{$5.65\cdot 10^{-6}$} & 
\multicolumn{3}{l|}{$5.65\cdot 10^{-7}$} & \multicolumn{3}{l|}{$5.65\cdot 
10^{-8}$} \\ \hline
$N$ &   50    &   60    &   70    &    50   &    60   &    70   &   50    &   
60 
   &    70   \\ \hline
 $r$ &   0.0341    &   0.0263    &   0.0209   &   0.0475    &   
0.0387    &   0.0325   &   0.0540    &   0.0447   &   0.0382    \\ 
\hline
$n_{s}$ &   0.9649    &   0.9700    &   0.9736    &   0.9674   &   
0.9728   &      0.9766   &   0.9678    &  0.9732   &   0.9769   \\ \hline
\end{tabular}
\vskip0.1cm
\begin{tabular}{|l|l|l|l|l|l|l|l|l|l|}
\hline
\multicolumn{10}{|l|}{$\alpha=2$, \ \ \ \ $n=4$, \ \ \ \ $V_{0}= 
10^{-16} M_{pl}^4$ 
  }\\ \hline
 $M$ &  \multicolumn{3}{l|}{$10^{-6}$} & \multicolumn{3}{l|}{$5 \cdot 10^{-7}$} 
 &\multicolumn{3}{l|}{$10^{-7}$} \\ \hline
 $\lambda$ &  \multicolumn{3}{l|}{$5.65\cdot 10^{-6}$} & 
\multicolumn{3}{l|}{$1.13\cdot 10^{-5}$} & \multicolumn{3}{l|}{$5.65\cdot 
10^{-5}$} \\ \hline
$N$ &   50    &   60    &   70    &    50   &    60   &    70   &   50    &   
60 
   &    70   \\ \hline
 $r$ &   0.0341    &   0.0263    &   0.0209    &  0.0415    &  0.0332 
   &  0.0273  &  0.0504    &   
0.0414    &     0.0351    \\ \hline
$n_{s}$ &   0.9649   &   0.9700    &   0.9736    &    0.9667    &  0.9720
   &   0.9757  & 0.9678   &   
0.9731   &      0.9769   \\ \hline
\end{tabular}
\caption{  Predictions for the   scalar spectral index  $n_\mathrm{s}$ and 
the tensor-to-scalar ratio $r$ for the scenario of non-canonical inflation with 
deformed-steepness potential, for fixed $\alpha=2$, $n=4$, for e-folding 
number $N$ being  
50, 60 and 70, and with fixed $M$ and varying $V_0$ (upper sub-Table) and with 
fixed $V_0$ and varying $M$ (lower sub-Table). In all cases    the 
value of $\lambda$   is determined  through the observational constraint 
(\ref{obscon}), while $M_{pl}=10^{18}$ $GeV$.} 
\label{Table2} 
\end{table}

The main observation is that the predictions of the scenario lie well inside 
the 1$\sigma$ region of observational data.
Now, for fixed $M$, increasing $V_0$ leads to lower values of $r$ and
$n_{\mathrm{s}}$; moreover, the variation of $r$ is much
faster than that of $n_{\mathrm{s}}$. Additionally,
 for fixed $V_0$, increasing $M$ leads also  to lower values of $r$ and
$n_{\mathrm{s}}$; nevertheless the change in $n_{\mathrm{s}}$ is   strongly
affected by the change in $M$ (since $M$ appears in powers of four in the 
equations) that it can easily be led outside the observational contours. 
This similar tendency behavior was   expected, since in the scalar-field 
equation (\ref{KLGord}) the  two parameters  $M$ and $V_0$  appear
multiplied. Nevertheless, this 
is not a trivial result, since $M$ is related to the non-canonicality scale 
while $V_0$ to the potential scale.

From the above analysis we deduce that  non-canonical kinetic terms
combined with deformed-steepness potentials can provide inflationary 
predictions in very good agreement with observations, compared to simple 
non-canonical models 
\cite{Barenboim:2007ii,Tzirakis:2008qy,Franche:2009gk,Franche:2010yj,
Unnikrishnan:2012zu,Gwyn:2012ey,Easson:2012id,Zhang:2014dja,Gwyn:2014wna, 
Hossain:2014xha,Rezazadeh:2014fwa,Cespedes:2015jga,Stein:2016jja, 
Dimopoulos:2017zvq,Mohammadi:2018wfk,Naderi:2018kre, 
Kamenshchik:2018sig,Do:2020ler} as well as to canonical models with 
deformed-steepness potentials 
\cite{Geng:2015fla,Rezazadeh:2015dia,Geng:2017mic,Ahmad:2017itq,Agarwal:2017wxo,
Skugoreva:2019blk,Das:2019ixt,Lima:2019yyv}. An additional significant 
 advantage is that the above combination allows to achieve good predictions 
without the need to use unnaturally large values for $\alpha$ or $n$, or 
unnaturally tuned values for the non-canonicality and potential scales $M$ and 
$V_0$, as well as for the potential exponent $\lambda$. In particular, we see 
that 
$M$ and $V_0$ remain in reasonable sub-Planckian regions, with values 
that can be easily predicted and accepted from field theoretical point of 
view. This combination 
of observational efficiency and theoretical justification is a significant 
advantage of the scenario at hand.

\section{Conclusions}

In this work we revisited   inflation with non-canonical scalar fields by 
applying deformed-steepness exponential potentials.
Non-canonical kinetic terms  can arise straightforwardly
in models of supergravity and superstrings, while 
 exponential potentials have remarkable properties,  as they greatly facilitate 
slow roll and result to scaling behaviour at large scales.

As we have shown, the resulting scenario can lead to inflationary observables, 
and 
in particular to    scalar spectral index of the curvature 
perturbations $n_\mathrm{s}$ and   tensor-to-scalar ratio $r$, in remarkable 
agreement with the observations of Planck 2018, being well inside the 1$\sigma$ 
region.
Apart from observational predictability, a significant additional 
advantage of the proposed scenario arises from the theoretical point of view. 
In 
particular, in order to obtain acceptable observables, in simple non-canonical 
models one needs to use relatively large non-canonical exponent $\alpha$ or  
ranges of values for the non-canonicality scale $M$, while in canonical models 
with deformed-steepness  potentials relatively large values of the extra 
exponent $n$ need to be imposed, and, hence, these models cannot be 
well-justified 
 theoretically. On the other hand, in the scenario of the present work the 
exponents $\alpha$ and $n$ are small, as well as the non-canonicality and 
potential scales
$M$ and $V_0$ remain in reasonable sub-Planckian regions.

We mention here that the motivation for the deformed-steepness 
parameter $n$ of \cite{Geng:2017mic} was the improved inflationary observables 
(specifically $r$)  compared to the   simplest slow-roll models. Moreover, the 
motivation for the  extra parameter $\alpha$ compared to canonical models is to  also
  improve $r$. Nevertheless, in works with non-minimal kinetic terms 
but with no  deformed-steepness parameter ($n=1$) the improvement of the 
observables is obtained for very large $\alpha$ values (of the order of 100 or 
1000 \cite{Unnikrishnan:2012zu}). Hence, the motivation for allowing both $n\neq1$ and $\alpha\neq1$  
is to be able to obtain improved observables, but with very low, and thus closer 
to natural, $\alpha$. Although the scenario has extra parameters, the improvement in observables (smaller tensor-to-scalar ratio values), as well as the higher naturalness is  worth the price.

Our analysis revealed that, for   a given $\alpha$, as $n$ increases both 
$n_{\mathrm{s}}$ and   $r$ decrease too, while on the other hand, for a given 
$n$, as $\alpha$ increases there is no particular tendency for 
$n_{\mathrm{s}}$ and   $r$.  
Additionally,   for fixed $M$, increasing $V_0$ leads to lower values of $r$ 
and $n_{\mathrm{s}}$, while 
 for fixed $V_0$, increasing $M$   leads  to lower values of $r$ and
$n_{\mathrm{s}}$ too.

In summary, we showed that revisiting non-canonical inflation models by 
applying potentials with deformed steepness, increases the observational 
predictability, bringing the  scalar spectral index and the tensor-to-scalar 
ratio more deeply into the observational contours, offering a better 
theoretical 
justification for the required parameters. This combination of observational 
efficiency and theoretical justification is a significant 
advantage of the scenario at hand, and hence,  non-canonical models with 
deformed-steepness potential need to be further explored, as
additional observational data will be coming forward.


\providecommand{\href}[2]{#2}\begingroup\raggedright\endgroup
\end{document}